\documentclass[3p, review]{elsarticle}

\usepackage{lineno,hyperref}
\usepackage{natbib,hyperref}
\usepackage{color}
\usepackage{amsmath, amssymb}
\usepackage{nomencl}
\usepackage{multicol}
\usepackage{url, arydshln}

\journal{}

\begin{document}

\begin{frontmatter}

\title{Performance analysis of a direct-absorption parabolic trough solar collector}

\author{Caiyan Qin}
\author{Joong Bae Kim}
\author{Bong Jae Lee\corref{cor}}

\cortext[cor]{Corresponding author}
\ead{bongjae.lee@kaist.ac.kr}

\address{Department of Mechanical Engineering, Korea Advanced Institute of Science and Technology, Daejeon 34141, South Korea}

\begin{abstract}
\color{black}A parabolic trough solar collector is a dominant technology for high-temperature industrial applications, but efficient use of a conventional surface-based parabolic trough solar collector (SBPTSC) is limited by its high radiation loss due to the high surface temperature. Recently, direct-absorption parabolic trough solar collector (DAPTSC) using nanofluids has been proposed, and its thermal efficiency has been reported to be 5--10\% higher than the conventional SBPTSC for inlet temperature up to 250$^\circ$C. However, the inner tubes of the receivers of the existing DAPTSCs are all transparent, so the sun rays entering the inner tube can only travel once through the nanofluids. As a result, the optical path length for the sun rays is limited by the inner tube size, which in turn requires high value of the absorption coefficient of nanofluids. Due to the approximately linear relation between the absorption coefficient and the particle concentration, higher absorption coefficient is likely to cause particle agglomeration, leading to detrimental effects on maintaining stable collector performance. In the current study, the transparent DAPTSC is improved by applying a reflective coating on the upper half of the inner tube outer surface, such that the optical path length is doubled compared to the transparent DAPTSC; thus, the absorption coefficient of the nanofluids can be reduced accordingly. The coated DAPTSC is found to have obvious advantage compared to the transparent DAPTSC at absorption coefficient below 0.5 cm$^{-1}$ for a receiver with inner tube diameter of 7 cm. In addition, performance of the transparent DAPTSC, the coated DAPTSC and the SBPTSC with black chrome coating have been compared to explore their advantageous operation conditions, such as inner tube diameter, flow rate, and inlet temperature, with or without a glass envelope for vacuum evacuation. The findings in this study will facilitate the use of nanofluids in a parabolic trough solar collector with more stability and provide a guide for choosing suitable type of parabolic trough solar collectors for specific working conditions.
\end{abstract}

\begin{keyword}
Direct-absorption solar collector \sep Parabolic trough solar collector \sep Blended plasmonic nanofluids
\end{keyword}

\end{frontmatter}

\section{Introduction}
Solar energy is a promising renewable energy in both aspects of sustaining the economic growth and reducing pollutants caused by energy consumption. Among the several available technologies for solar energy harvesting, the concentrated solar power (CSP) is an emerging and significant technology with the advantages, such as built-in storage capability and low green house gas emissions \cite{yilmaz2018modeling}. The parabolic trough solar collector is a dominant CSP technology among the medium-high-temperature solar collectors working for the operating temperature up to 800 K \cite{kalogirou2013solar}. In the parabolic trough solar collector, a mirror or a reflector with parabola shape concentrates the solar radiation to its focal line where a tubular receiver containing working fluid is mounted to generate high-temperature heat. So far, the receiver is usually composed of a coated metal absorber (e.g., black chrome coating) and a glass envelope, and the annulus between them is evacuated to suppress the convection heat loss. The desirable absorber surface should have high absorptivity in the solar spectrum but low emissivity in the mid-infrared spectrum to minimize the radiation loss. The major disadvantages of the surface-based parabolic trough solar collector (SBPTSC) are the low thermal efficiency at high temperature \cite{fan2018heat} and the instability of selective coating at high temperatures \cite{freedman2018analysis}. To remedy those issues associated with the SBPTSC, a direct-absorption parabolic trough solar collector (DAPTSC) using nanofluids has been proposed recently, and its thermal efficiency has been found 5--10\% higher than the SBPTSC \cite{khullar2012solar}. Therefore, the DAPTSC can be an efficient alternative for harvesting solar energy at high temperature.

Nanofluids is a dispersion of nanoparticles in a base fluid, such as water or oil \cite{tyagi2009predicted}. Due to the capability of nanoparticles that can directly interact with the solar radiation, nanofluids can absorb the solar radiation uniformly inside. As a result, the temperature distribution inside the solar collector becomes relatively uniform, leading to less heat loss from the collector surface and improved thermal efficiency \cite{tyagi2009predicted, otanicar2010nanofluid}. Nanofluids has been studied intensively for direct absorption solar collectors (DASCs) in low-temperature applications \cite{sani2011potential, lee2012radiative, veeraragavan2012analytical, gupta2015investigation, cregan2015modelling, gorji2015geometry, jeon2016theoretical, qin2017optimization, won2018effect, qin2018optimization, mallah2018blended}. In addition, the optical property of various nanofluids have also been studied intensively to improve its photothermal conversion efficiency \cite{jeon2014optical, chen2015anexperimental, chen2016enhancement, du2016plasmonic, bhalla2018parameters, wang2018numerical}.

On the basis of the study on DASCs using nanofluids, a direct-absorption parabolic trough solar collector (DAPTSC) for high temperature applications has been studied recently. Khullar et al. \cite{khullar2012solar} first proposed the idea of utilizing aluminum nanofluids with a volume fraction of 5\% for concentrating solar collector, and their DAPTSC was reported to have about 5--10\% higher thermal efficiency than the conventional SBPTSC. Later, several numerical works have been performed to investigate the effect of nanofluid concentrations, mass flow rate and inlet temperature for the DAPTSC \cite{menbari2016heat, o2018modelling2, o2018modelling1, freedman2018analysis}. However, all these studies employed the transparent glass inner tube; thus, the sun rays can only travel once through the nanofluids after entering the inner tube. Consequently, the optical path length for sun rays is limited by the inner diameter of the inner tube of the receiver, which in turn requires high value of the absorption coefficient of nanofluids to gain sufficiently high energy absorption \cite{jeon2016theoretical}. As low absorption coefficient indicates low particle concentration \cite{bohren1983absorption}, which is beneficial for avoiding particle agglomeration, a DAPTSC should be designed to have a larger diameter of inner tube to reduce the particle concentration of nanofluids. Besides, all these studies considered nanofluids with a spectrally varying absorption coefficient simply depending on nanoparticle material (i.e., uncontrolled manner). However, Qin et al. \cite{qin2018optimization} recently showed that the blended plasmonic nanofluid with either a uniform absorption coefficient (when the nanoparticle concentration is sufficiently high) or a spectrally varying absorption coefficient following the shape of solar spectrum (when the nanoparticle concentration is below a certain value) is more beneficial for broadband solar energy absorption. In addition, a uniform solar irradiance is assumed in recent works \cite{khullar2012solar, de2013modelling, fan2018heat}, which is not practical in case of parabolic trough solar collectors where the solar radiation is highly dependent on the angular direction of the receiver. Lastly, in Refs. \cite{khullar2012solar, o2018modelling2}, the radiation from the inner tube surface is assumed to be dissipated to the environment directly, but this assumption is not valid because the glass envelope is not transparent to the long-wavelength thermal emission in the mid-infrared spectral region \cite{incropera2011principles}.

The present work carries out the performance analysis of a new type of direct-absorption parabolic trough solar collector. Inspired by the flat-plate DASCs where a mirror is employed at the bottom of the channel to enhance the optical path length of sun rays inside the collector \cite{lee2012radiative, gorji2016numerical}, we propose to apply a highly reflective metallic coating on the upper half of the inner tube outer surface such that the optical path length for the concentrated rays can be doubled due to the reflection. It is worth mention that the radiation heat loss from the inner tube outer surface can also be reduced greatly due to the low emissivity of a metallic coating compared to that of bare glass. Besides, the blended plasmonic nanofluid with uniform absorption coefficient will be applied to have broadband absorption of solar energy. In addition, the ray tracing method will also be employed in this study to obtain the nonuniform solar irradiance on the inner tube so that the corresponding nonuniform heat generation by the blended plasmonic nanofluid can be obtained.

\section{Theoretical model of a direct-absorption parabolic trough solar collector}
In order to simulate the DAPTSC, two models were developed in COMSOL Multiphysics\textsuperscript{\textregistered}. First, a two-dimension (2-D) optical model was built to simulate the solar incidence on the collector based on the ray tracing method as well as the corresponding heat source generation based on the Beer-Lambert law. Second, a three-dimensional (3-D) thermal model that couples the turbulent fluid flow with the heat transfer inside the inner tube of the receiver was built. The heat source obtained from the 2-D optical model was incorporated to the 3-D thermal model for calculating the velocity and the temperature distributions inside the inner tube of the receiver. Because there is no difference along the tube length in the volumetric heat generation inside the nanofluid, the 2-D optical model is sufficient for calculating the heat source for the 3-D thermal model, leading to substantially reduced calculation time.

\begin{figure}[!t]
\centering
\includegraphics[width=.6\textwidth]{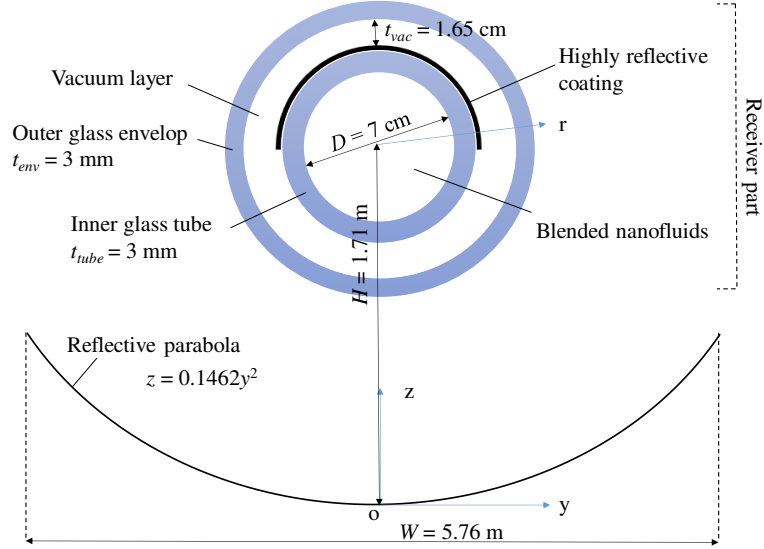}
\caption{Schematic of the coated DAPTSC.}
\label{fig1}
\end{figure}

The schematic of the 2-D model is shown in Fig.\ \ref{fig1} with geometrical sizes annotated. Sizes of the parabola and the receiver are chosen according to the commercialized LS-3 trough and the LS-3 receiver, respectively \cite{cheng2014comparative}, and its geometrical concentration ratio is $W/D\approx 82.3$. In this study, only the normally incident solar radiation is considered for simplicity. The parabolic trough solar collector system consists of a receiver part (two concentric glass tubes, i.e., an inner glass tube and an outer glass envelope) and one aluminum parabola with a high reflectivity of 0.94 \cite{cheng2014comparative, fan2018heat}. The parabola is used to reflect and concentrate the incident sun rays to the receiver tubes. The glass envelope is transparent in the solar spectrum with the transmittance of $\tau_{gla}=0.94$ \cite{khullar2012solar, cheng2014comparative, fan2018heat} but has emissivity of $\epsilon_{gla}=0.86$ in the mid-infrared spectral region \cite{li2016experimental, fan2018heat}. The property of the inner tube is the same with the envelope glass, except that a highly reflective metallic coating is applied on its upper half of the outer surface in order to double the optical path length of the concentrated rays. Notice that among all the rays incident onto the coated DAPTSC, only a small portion ($D/W\approx1.2\%$) are blocked by the metallic coating on the outer surface of the inner tube while most other rays are intercepted by the parabola reflector and then concentrated to the receiver. Therefore, compared to the transparent DAPTSC, about 1.2\% of the solar radiation is blocked in order to gain the doubled optical path length in the coated DAPTSC. Another advantage of the reflective coating is that due to the metal's much lower infrared emissivity than that of glass \cite{incropera2011principles}, it can significantly reduce the radiation loss from the inner tube outer surface.

In the evaluation of the extinction of the incident rays, the scattering effect is ignored by assuming the small size of suspended nanoparticles, usually several tens of nanometers, as also done in other studies related to DASC with nanofluids \cite{tyagi2009predicted, otanicar2010nanofluid, veeraragavan2012analytical, jeon2014optical, rativa2015solar, cregan2015modelling}. In the current study, the absorption coefficient $\alpha$ of the blended plasmonic nanofluid is assumed to be constant in the solar spectrum for achieving broadband solar energy absorption. This assumption is valid when nanoparticles with different absorption peaks are well mixed together \cite{jeon2016theoretical, qin2017optimization}. Note also that the magnitude of $\alpha$ is approximately proportional to the concentration of the nanoparticles and thus, can be tuned by varying the nanoparticle concentration \cite{lee2012radiative, qin2018optimization}. Therefore, the spectral uniformity and the magnitude of the absorption coefficient of nanofluid can be individually tuned. The solar irradiance reaching the inner tube surface is numerically calculated from the 2-D optical model using the ray tracing method. The attenuation of the solar radiation in the working fluid is taken into consideration according to the Beer-Lamber law, resulting in the locally varying volumetric heat generation, $\dot{Q}_{gen}(r, \phi)$. The solar weighted absorption coefficient, $A_m$, can be estimated from 
%
\begin{equation}
A_m=\frac{1}{WG_T}\int_0^{D/2}\int_0^{2\pi}\dot{Q}_{gen}(r, \phi)rdrd\phi
\end{equation}
%
where $G_T$ is the solar irradiance assumed as 992 W/m$^2$.

In the 3-D thermal model, the fluid flow and heat transfer are coupled in the inner tube region. The thermophysical property of the nanofluids are taken the same as the base fluids, i.e., Therminol VP-1 \cite{solutia2013}, because the nanoparticle concentrations is far below 1\% in volume fraction, as was also done in other related works \cite{khullar2012solar, lenert2012optimization, o2018modelling1, freedman2018analysis, dugaria2018modeling}. The flow rate in this study is around 0.1 kg/s and the corresponding Reynold number is above $\text{Re}_D = 2300$; thus, the flow is in the turbulent regime and the standard $k-\varepsilon$ model was applied with corresponding parameters $C_{\mu}=0.09$, $c_{\varepsilon 1}=1.44$, $c_{\varepsilon 2}=1.92$, $\sigma_k=1$ and $\sigma_\varepsilon=1.3$ \cite{launder1983numerical}. The wall function method was used in the vicinity of the solid wall because the viscosity effects predominate over the the turbulent effects \cite{launder1983numerical}. For thermal analysis, the following energy equation was solved \cite{bergman2011fundamentals}:
%
\begin{equation}
\frac{\partial}{\partial x_i}(\rho u_i T) = \frac{\partial}{\partial x_i} \left [ (k+k_t)\frac{T}{x_i} \right]+\dot{Q}_{gen}
\end{equation}
%
where the subscript `$i$' represents the coordinate $x$, $y$ or $z$, and $u_i$ and $x_i$ indicates the velocity and dimension in the corresponding directions respectively. In the above equation, $\rho$ is the fluid density, $k$ is the fluid thermal conductivity, $k_t$ is the turbulent thermal conductivity defined as $\mu_t C_p/Pr_t$ with turbulent viscosity $\mu_t$ from the turbulent flow model and turbulent Prandtl number from the Kays-Crawford model \cite{kays1994turbulent}, and $\dot{Q}_{gen}$ is transformed to the Cartesian coordinate. The boundary conditions are:
%
\begin{equation}\begin{gathered}
T(x=0)=T_0\\ 
\frac{\partial T}{\partial x} \biggm |_{x=L}=0
\end{gathered}\end{equation}
%
where $T_0$ is the inlet temperature and $L$ is the length of the receiver tube. And 
\begin{equation}
- (k+k_t)\frac{\partial T}{\partial r} \biggm |_{r=R_{to}} = \frac{\sigma \left [\bar{T}^4_{ei}-T^4(r=R_{to}) \right ]}{\frac{2}{\epsilon_{to,u}+\epsilon_{to,l}}+\frac{1-\epsilon_{ei}}{\epsilon_{ei}} \left (\frac{D_{to}}{D_{ei}} \right )} + h_{in} \left [\bar{T}_{ei}-T(r=R_{to}) \right]
\label{Eq_BC2}
\end{equation}
%
where the subscripts `$ei$' denotes the envelope inner surface, `$to,u$' and `$to,l$' denotes the upper half and lower half of the inner tube outer surface, respectively. In the above equation, the surface emissivity values are $\epsilon_{ro,u}=0.1$, $\epsilon_{ro,l}=0.86$ and $\epsilon_{ei}=0.86$, $\bar{T}_{ei}$ denotes the average of the envelope inner surface temperature, and $h_{in}$ is the convective heat transfer coefficient in the evacuated annulus which is assumed to be $h_{in}=0.000174$ W/m$^2$-K \cite{forristall2003heat}. The radiation term in Eq.\ (\ref{Eq_BC2}) is derived based on the radiation network approach \cite{incropera2011principles} assuming that the whole inner tube surface is at its average temperature. It is also recognized that for using the radiation network approach, the surface is assumed as diffusive and gray. Even though the glass surface is not real diffusive, it is an easy and good approximation by assuming diffusive as also done in other studies \cite{xu2015performance, yilmaz2018modeling}.

Note that some previous works \cite{khullar2012solar, o2018modelling2} assumed that the radiation heat loss from the inner tube is dissipated directly to the environment through the glass envelope. Unfortunately, this simple assumption is not valid because that glass envelope is not transparent to the infrared thermal radiation \cite{incropera2011principles}. In the current study, the heat loss from the inner tube is treated to be transferred to the envelope inner surface first and then this heat dissipates to the ambient via heat convection as well as to the space via thermal radiation. In order to obtain $\bar{T}_{ei}$, the energy balance equation is applied again to the glass envelope; that is, the heat transfer from the outer surface of the inner tube to the inner surface of the envelope should be the same as the heat transfer from the outer surface of the envelope to the surroundings:
%
\begin{equation}\begin{gathered}
\left ( \frac{D_{ei}}{D_{to}} \right ) \left[{\frac{\sigma \left [ T^4(r=R_{ei})-\bar{T}^4_{to} \right]}{\frac{2}{\epsilon_{to,u}+\epsilon_{to,l}}+\frac{1-\epsilon_{ei}}{\epsilon_{ei}} \left ( \frac{D_{to}}{D_{ei}} \right )}}+h_{in} \left [T(r=R_{ei})-\bar{T}_{to}) \right] \right] = \\
\left ( \frac{D_{eo}}{D_{ei}} \right) h_{out} \left [ T_{a}-T(r=R_{eo}) \right ]+ \epsilon_{eo}\sigma \left [T_{sky}^4-T^4(r=R_{eo}) \right]
\label{Eq_BC3}
\end{gathered}\end{equation}
%
where the subscripts `$eo$' denotes the envelope outer surface and $\bar{T}_{to}$ is the average of inner tube outer surface temperature, $h_{out}$ is natural heat transfer coefficient at the glass envelope outer surface which is estimated as 10 W/m$^2$-K \cite{wang2016parabolic}. $T_a$ is ambient temperature with 293.15 K and  $T_{sky} $ is sky temperature assumed as 278 K \cite{swinbank1963long}. Note that in Eq.\ (\ref{Eq_BC3}) the term $(\frac{D_{ei}}{D_{to}})$ and $(\frac{D_{eo}}{D_{ei}})$ are used to guarantee the heat flux balance. Finally, the collector thermal efficiency is defined as \cite{duffie2013solar}:
%
\begin{equation}
\eta = \frac{\dot{m} c_p (\bar{T}_{out}-T_0)}{LW G_T}
\label{Eq_eta}
\end{equation}
where $\dot{m}$ is the flow rate, $\bar{T}_{out}$ is the average temperature at the outlet, and $c_p$ is the heat capacity at the average temperature, i.e., $T_m$ = $(T_0+\bar{T}_{out})/2$. In addition, the temperature gain of the working fluid through the collector is estimated from $\Delta T=\bar{T}_{out}-T_0$.

In order to validate the thermal modeling, the calculation results of the SBPTSC are compared to the experimental results from Sandia National Laboratory \cite{dudley1994test} and the predicted results by Xu et al. \cite{xu2015performance} for several selected cases, using the same parameters for configuration (LS-2 trough and collector) and the same base fluids (Syltherm 800). Note that in case of the SBPTSC, the inner tube outer surface absorbs the solar radiation directly, and thus, the surface heat source term is obtained from the 2-D optical model and is directly assigned to the 3-D thermal model. The test results are listed in Table \ref{Tab1}. It is seen that the current simulation results are generally in between of the experimental results and the numerical simulation results. The reason that the current thermal efficiency is slightly higher than the experimental data is probably because the current model did not account for the heat loss through the brackets and tube ends. Because the relative difference of the current results with the experimental data and other numerical results are within 5\%, the current thermal model can be considered to be reliable. 

\begin{table}[!t]
\centering
\caption{Comparison of experimental and numerical data with the current simulation results. The unit for ambient temperature $T_a$, inlet temperature $T_0$ and temperature gain $\Delta T$ are all in K, the unit for solar irradiance $G_T$ is W/m$^2$, and that for volumetric flow rate $V_{in}$ is L/min. The subscripts of `$exp$', `$num$' and `$cur$' indicates the experimental results \cite{dudley1994test}, numerical results \cite{xu2015performance} and the current simulation results, respectively.}
\label{Tab1}
\begin{tabular}{ c || c  c  c  c | c  c :  c  | c  c : c  }
\hline
 & $G_T$ & $T_a$ & $T_0$ & $V_{in}$ & $\Delta T_{exp}$ & $\Delta T_{num}$ & $\Delta T_{cur}$ & $\eta_{exp}$ & $\eta_{num}$ & $\eta_{cur}$ \\ \hline
Case 1 & 933.7 &294.35&375.35&47.7&21.8 &22.40 &22.55 &0.7251 &0.7424 &0.7440 \\
Case 2 & 968.2 &295.55&424.25&47.8&22.3 &23.21 &23.01 &0.7090 &0.7394 &0.7279 \\
Case 3 & 982.3 &297.45&470.65&49.1&22.0 &22.94 &22.68 &0.7017 &0.7310 &0.7154 \\ \hline
\end{tabular}
\end{table}

\begin{figure}[!b]
\centering
\includegraphics[width=.5\textwidth]{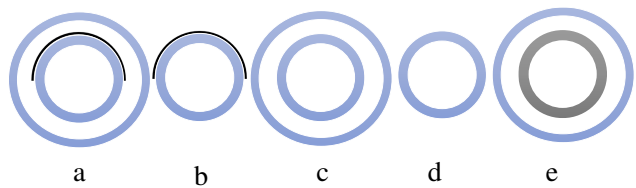}
\caption{Schematics of the receiver part of the five solar collectors: (a) coated DAPTSC with a glass envelope; (b) coated DAPTSC without a glass envelope; (c) transparent DAPTSC with a glass envelope; (d) transparent DAPTSC without a glass envelope; (e) conventional SBPTSC with a glass envelope.}
\label{fig2}
\end{figure}

In this study, the other four collectors, including the coated DAPTSC without a glass envelope, the transparent DAPTSC, the transparent DAPTSC without a glass envelope, and the conventional SBPTSC with black chrome coating, are investigated and compared with each others (refer to Fig.\ \ref{fig2}). Because their modeling approach are generally similar with that for the coated DAPTSC, only the difference are briefly described as below. For the transparent DAPTSC, the only difference is that the the term $\frac{2}{\epsilon_{to,u}+\epsilon_{to,l}}$ is replaced by $\frac{1}{\epsilon_{to}}$ in Eqs.\ (\ref{Eq_BC2}) and (\ref{Eq_BC3}) as the whole inner tube outer surface has the same emissivity of 0.86. As to the SBPTSC, besides the above mentioned difference, the inner tube surface absorbs the solar energy directly (with an absorptance of 0.94) to obtain the boundary heat source term for the 3-D model. As for the coated DAPTSC as well as the transparent DAPTSC without a glass envelope, the modeling approach is similar as their counterparts with a glass envelope, except that the calculation for the 3-D model is much simpler since the inner tube surface is directly facing the environment. Note also that for SBPTSC, its inner tube (or absorber) surface has a temperature-dependent thermal emissivity $\epsilon_{m}$ in the range of $0.2 \sim 0.4$ assuming the surface as black chrome coated copper \cite{li2016experimental,o2018modelling1}.

%

\section{Results and discussion}

\subsection{Effects of the absorption coefficient of the blended plasmonic nanofluid}
Absorption coefficient $\alpha$ of the blended plasmonic nanofluid is an important parameter for the thermal performance of DAPTSCs by influencing the local heat generation inside the inner tube. In this section, effects of the absorption coefficient are investigated for the transparent DAPTSC and the coated DAPTSC, both with a glass envelope. The configuration of the two collectors are all the same except that a highly reflective metallic coating is applied on the upper half of the inner tube outer surface for the coated DAPTSC [see Fig.\ \ref{fig2}(a)] while no such a coating exists for the transparent DAPTSC [see Fig.\ \ref{fig2}(c)]. The inlet temperature $T_0$ is set to be 550 K, which is a typical working temperature for parabolic trough solar collector \cite{khullar2012solar, freedman2018analysis}, the flow rate $\dot{m}$ is chosen as 0.1 kg/s, which is comparable to other studies \cite{li2016experimental, dugaria2018modeling}, the inner tube inner diameter $D$ is chosen as 7 cm (according to the LS-3 receiver), and the collector length $L$ is chosen as 5 m for all the evaluation.
\begin{figure}[!t]
\centering
\includegraphics[width=0.9\textwidth]{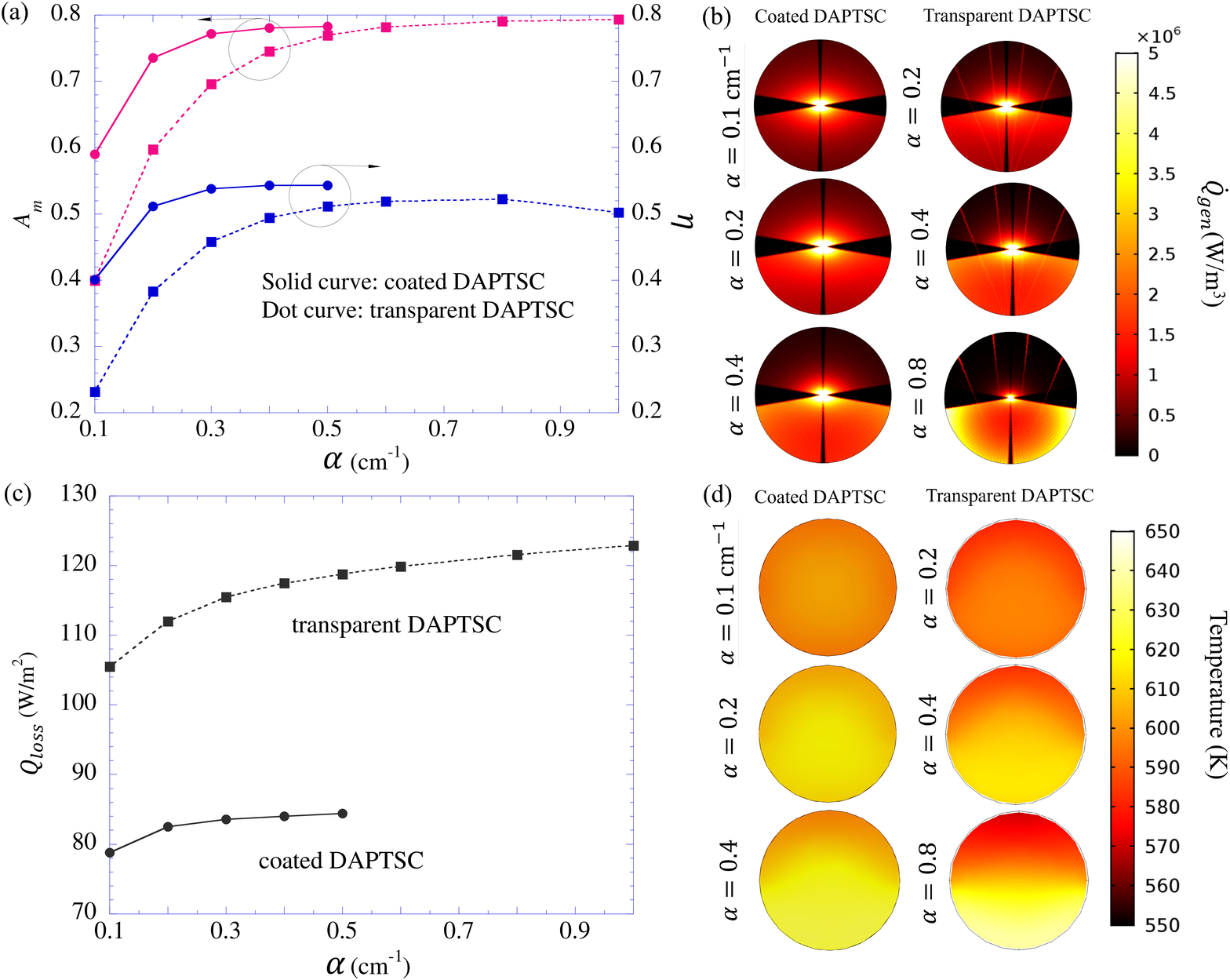}
\caption{Effect of the absorption coefficient on: (a) solar weighted absorption coefficient $A_m$ and collector thermal efficiency $\eta$; (b) volumetric heat generation in the nanofluid; (c) heat loss from the receiver; and (d) temperature distribution at the the tube outlet. In the calculation, we set $T_0$ = 550 K, $D$ = 7 cm, $L$ = 5 m, and $\dot{m}$ = 0.1 kg/s.}
\label{fig3}
\end{figure}

It can be seen from Fig.\ \ref{fig3}(a) that the solar weighted absorption coefficient $A_m$ keeps increasing with increasing $\alpha$ for both coated and transparent DAPTSCs. The increasing rate gets slower with increasing $\alpha$ and becomes zero near $\alpha$ = 0.4 cm$^{-1}$ for the coated DAPTSC and near $\alpha$ = 0.8 cm$^{-1}$ for the transparent DAPTSC. Note that $1-e^{-0.4\times2D} \text{ (for the coated DAPTSC)} = 1-e^{-0.8\times D} \text{ (for the transparent DAPTSC)}=0.996$, indicating that 99.6\% of the solar energy entering the inner tube is absorbed by the nanofluid, which is also reflected in Fig.\ \ref{fig3}(b). For instance, when $\alpha=0.8$ cm$^{-1}$ for the transparent DAPTSC, the heat source in the upper half of the inner tube is quite low because almost all the energy has been absorbed at the lower part. It is interesting to see that the maximum value of $A_m$ for the coated DAPTSC (i.e., 0.783) is slightly lower than that for the transparent DAPTSC (i.e., 0.793). This is because about 1\% (i.e., $D/W \approx 0.01$) of the solar radiation is blocked by the reflective metal coating for the coated DAPTSC. 

Similarly, the collector thermal efficiency $\eta$ also increases with the increase of $\alpha$ at first and then starts to decrease when exceeding certain $\alpha$ value [e.g, about 0.4 cm$^{-1}$ for the coated DAPTSC or about 0.8 cm$^{-1}$ for transparent DAPTSC; refer to Fig.\ \ref{fig3}(a)]. The reason is that, at lower value of $\alpha$, the increase of $\alpha$ causes more solar energy to be absorbed in the nanofluid as represented by $A_m$ in Fig.\ \ref{fig3}(a). However, when $\alpha> 0.4$ cm $^{-1}$ for the coated DAPTSC or $\alpha> 0.8$ cm $^{-1}$ for the transparent DAPTSC, the contribution by increasing $\alpha$ to $A_m$ becomes limited while the heat loss keeps increasing, as shown in Fig.\ \ref{fig3}(c). Here, $Q_{loss}$ is defined as the total heat loss (i.e., both heat convection and thermal radiation from the envelope outer surface) divided by the interception area of the parabola (i.e., $W\times L$). The increased heat loss to the surrounding eventually leads to the decrease in the collector efficiency at those conditions. 

It can be also seen from Fig.\ \ref{fig3}(a) that at the same absorption coefficient $\alpha$, the coated DAPTSC performs better than the transparent DAPTSC, especially for lower values of $\alpha$ region. At $\alpha=0.2$ cm$^{-1}$, for instance, $A_m$ and $\eta$ of the coated DAPTSC are much higher than those for the transparent DAPTSC. In other words, the coated DAPTSC requires lower value of $\alpha$, thus lower nanoparticle concentration \cite{bohren1983absorption} than the transparent DAPTSC for the same level of the thermal performance, which is beneficial for avoiding particle agglomeration and maintaining stable operation. If $\alpha > 0.5$ cm$^{-1}$, then the advantage of the coated DAPTSC becomes minimal because the absorption of the solar radiation is already high enough (i.e., $1-e^{-0.5\times D}=0.969$) for the transparent DAPTSC. Nevertheless, due to the low emissivity of the metallic reflective coating, the coated DAPTSC still exhibits the thermal efficiency of about 0.551, which is slightly higher than $\eta=0.520$ of the transparent DAPTSC.

It is important to note that the transparent DAPTSC with doubled $\alpha$ values of the coated DAPTSC (i.e., both collectors have the same value of `$\alpha \times \text{optical path length}$') cannot perform the same as the coated DAPTSC. For instance, $\eta$ of the transparent DAPTSC with $\alpha$ = 0.4 and 0.8 cm$^{-1}$ is lower than $\eta$ of the coated DAPTSC with $\alpha$ = 0.2 and 0.4 cm$^{-1}$, respectively [refer to Fig.\ \ref{fig3}(a)]. The main reason is that the emissivity of the inner tube surface for the coated DAPTSC (i.e., half glass surface and half metallic coating) is lower than that of the transparent DAPTSC (whole glass surface), leading to lower radiation loss. Another reason is that when $\alpha$ is doubled for the transparent DAPTSC, the heat source as well as the resulting temperature distribution will be not as uniform as the coated DAPTSC at the corresponding $\alpha$ values (i.e., half of the transparent DAPTSC), leading to locally high temperature on the inner tube surface thus more heat loss [refer to Figs.\ \ref{fig3}(b) and \ \ref{fig3}(d)].

\subsection{Effects of the inner diameter of the inner tube}
In this section, effects of the inner diameter of the inner tube, $D$, is investigated, and comparisons are made for the coated DAPTSC, the transparent DAPTSC and the SBPTSC with black chrome coating, all with a glass envelope. For fair comparison, we set $\alpha D$ to be constant for different values of $D$ for nanofluid-based DAPTSCs. Here, when $D$ is varied, the envelope diameter also changes accordingly to make the annulus spacing unchanged. 
\begin{figure}[!b]
\centering
\includegraphics[width=0.9\textwidth]{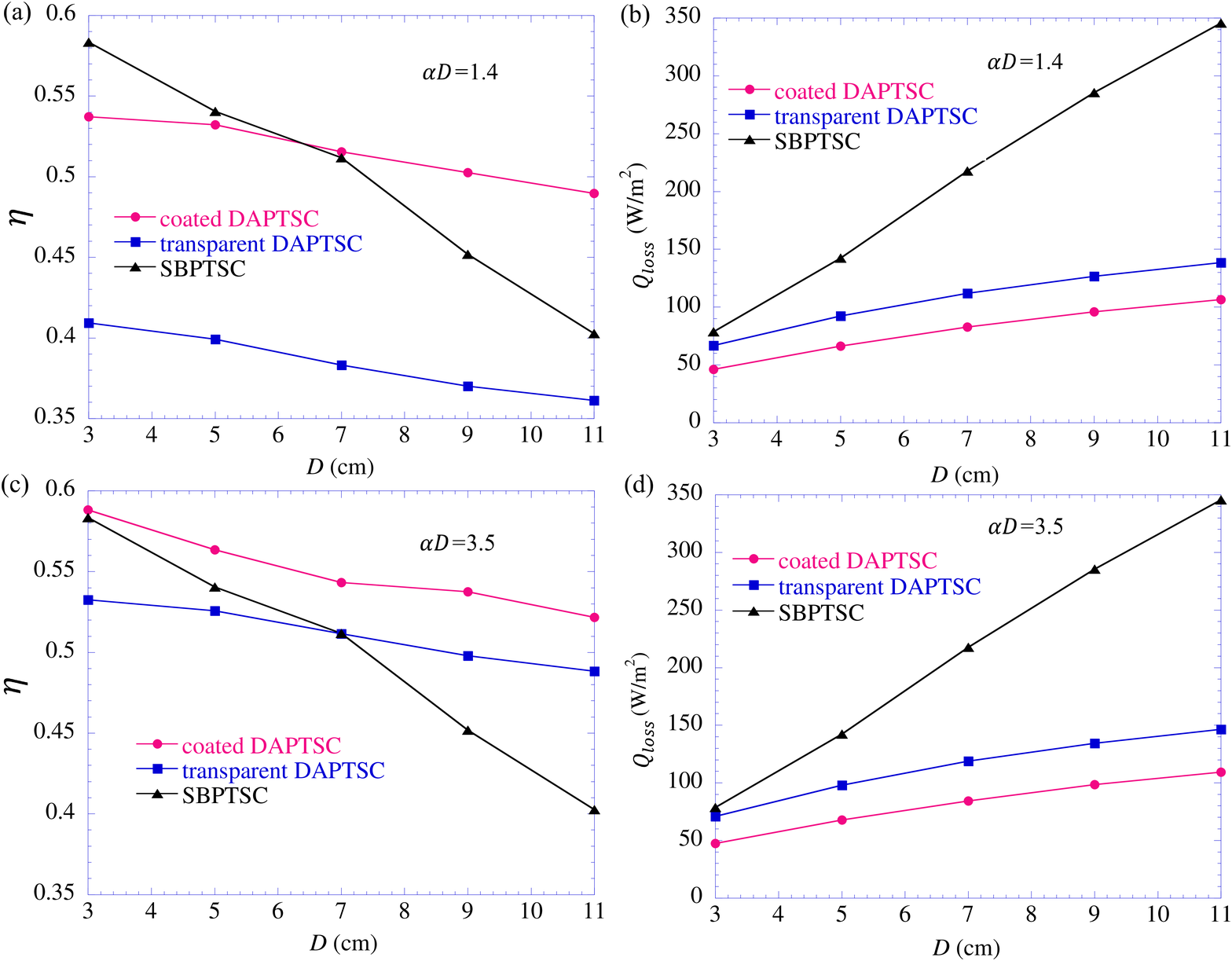}
\caption{Effect of the tube inner diameter on: (a) and (c) collector thermal efficiency; and (b) and (d) heat loss from the receiver. In the calculation, we set $T_0$ = 550 K, $\dot{m}$ = 0.1 kg/s, and $L$ = 5 m. Note that in (a) and (b), $1-e^{-\alpha D}= 0.753$, and in (c) and (d), $1-e^{-\alpha D}= 0.969$.}
\label{fig4}
\end{figure}

Figures \ref{fig4}(a) and \ref{fig4}(b) show that $\eta$ of all the three collectors decreases with increasing $D$ due to nearly constant $A_m$  value (i.e., 0.735, 0.597 and 0.849 for the coated DAPTSC, the transparent DAPTSC and the SBPTSC, respectively) as well as increasing heat loss. Besides, the decrease rate of $\eta$ for DAPTSCs is much slower than the SBPTSC owing to their much lower heat loss than the SBPTSC, which arises from the fact that the inner tube surface temperature of the SBPTSC is much higher than the nanofluid-based DAPTSCs (not shown here). Figure \ref{fig4}(a) suggests that the DAPTSCs are more suitable for collectors with large diameter than the SBPTSC. For instance, the coated DAPTSC performs better than the SBPTSC when $D > 7$ cm. It should also be noted that for the given level of $\alpha D$ value, the DAPTSCs require smaller $\alpha$ (i.e., lower particle concentrations) with increasing $D$, which is beneficial for avoiding particle agglomeration and maintaining stable collector performance. Because $\alpha$ is inversely proportional to $D$ when $\alpha D = \text{const}$, $|\frac{d\alpha}{dD}|$ becomes lager when $D$ is smaller. In principle, larger $\eta$ and smaller $\alpha$ are desirable; thus, there exists an optimal value of $D$ where the DAPTSCs can gain a reasonably high $\eta$ at a sufficiently low $\alpha$ for stable operation. For instance, the coated DAPTSC with $\alpha$ = 0.39 cm$^{-1}$ in Fig.\ \ref{fig4}(a) has $\eta= 0.538$, which is more advantageous than the one obtaining slightly higher $\eta=0.543$ but with much higher value $\alpha=0.5$ cm$^{-1}$.

Now the value of $\alpha D$ increases from 1.4 to 3.5 in Figs.\ \ref{fig4}(c) and \ref{fig4}(d), along with the same SBPTSC as in Figs.\ \ref{fig4}(a) and \ref{fig4}(b) for comparison. We can see that the DAPTSCs with $\alpha D=3.5$ show higher $\eta$ values. This is mainly due to the increased $A_m$ (i.e., 0.783 and 0.769 for the coated DAPTSC and the transparent DAPTSC, respectively). Comparison of Fig.\ \ref{fig4}(a) with Fig.\ \ref{fig4}(c) reveals that for the nanofluid of the same $\alpha$, DAPTSCs with different $D$ values may achieve the same level of the performance $\eta$. For instance, when $\alpha= 0.35$ cm$^{-1}$, the coated DAPTSC with $D=4$ cm or $D=10$ cm can both result in $\eta \approx 0.53$. Even though $A_m$ for $D$ = 4 cm case in Fig.\ \ref{fig4}(a) is 0.74 and smaller than 0.78 for $D$ = 10 cm in Fig.\ \ref{fig4}(c), the corresponding heat loss is about 56 W/m$^2$ in Fig.\ \ref{fig4}(b) and much less than 102 W/m$^2$ in Fig.\ \ref{fig4}(d). Therefore, to obtain the same $\eta$ (or similar level of performance) with nanofluid of the same $\alpha$, there could be multiple possible $D$ values to choose.

\subsection{Effects of the flow rate}
\begin{figure}[!t]
\centering
\includegraphics[width=0.9\textwidth]{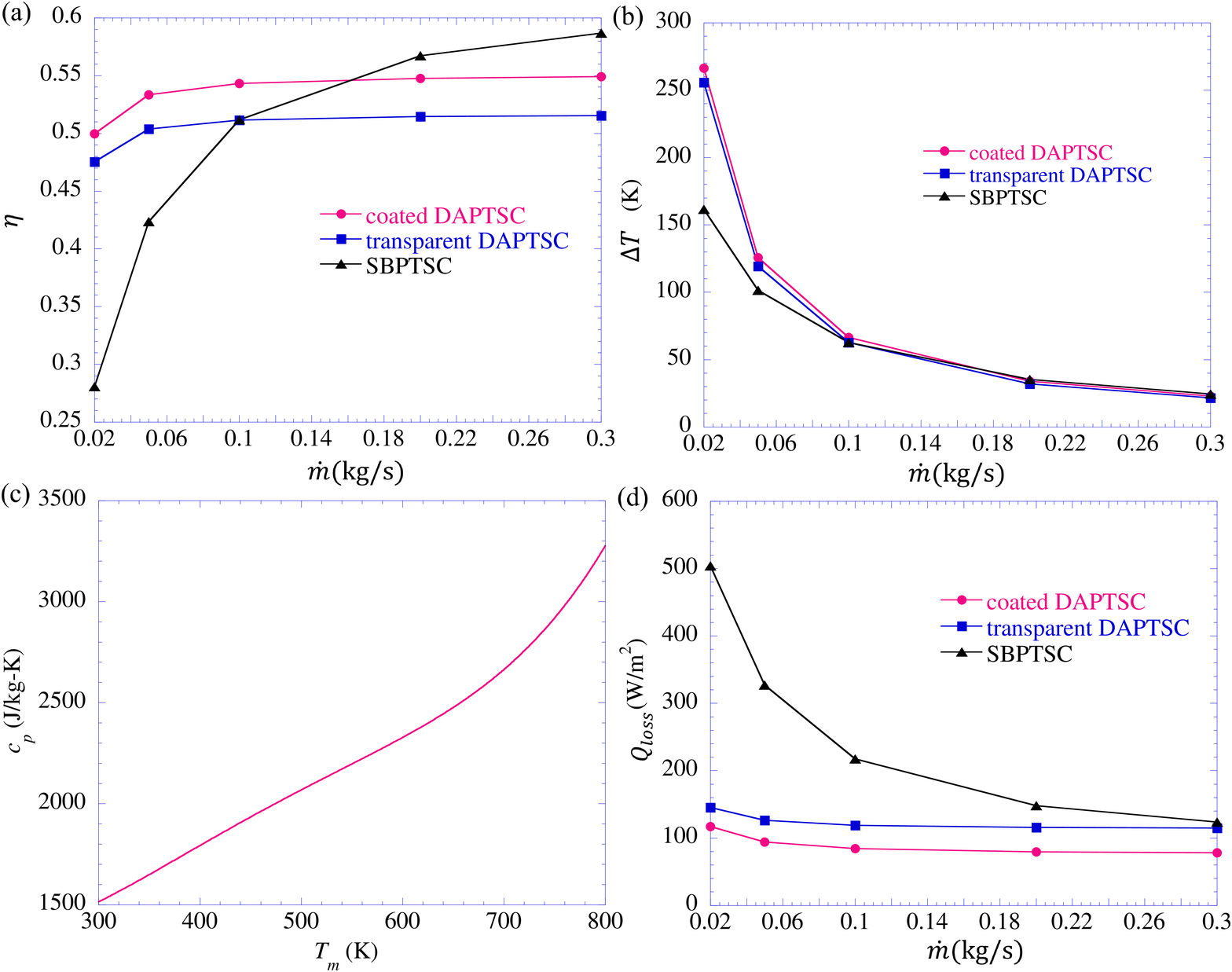}
\caption{Effects of the flow rate on: (a) thermal efficiency; (b) temperature gain; (c) temperature-dependent heat capacity of Therminol VP-1; and (d) heat loss. In the calculation, we set $\alpha$ = 0.5 cm$^{-1}$, $T_0$ = 550 K, $D$ = 7 cm, and $L$ = 5 m.}
\label{fig5}
\end{figure}

The effect of flow rate on the collector performance is investigated here in the range from 0.02 to 0.3 kg/s with the corresponding $\text{Re}_D = 600 \sim 9000$. Performance of three collectors including the coated DAPTSC, the transparent DAPTSC and the SBPTSC are compared in Fig.\ \ref{fig5}. It can be seen from Fig.\ \ref{fig5}(a) that as the flow rate $\dot{m}$ increases, the collector thermal efficiency $\eta$ increases but its increasing rate slows down gradually. This can be explained based on the following two perspectives. 

First, recall that $\eta$ is determined by Eq.\ (\ref{Eq_eta}) and any change in $\eta$ is due to the variation of the three parameters, i.e., $\dot{m}$, $\Delta T$ (or $\bar{T}_{out}-T_0$) and $c_p$. With the increase of $\dot{m}$, $\Delta T$ first decreases significantly and then the decrease rate becomes slower, as shown in Fig.\ \ref{fig5}(b). In addition, $c_p$ also decreases with slower decreasing rate as $\dot{m}$ increases because it has approximately linear relation with the average temperature $T_m$ (i.e., $\frac{T_0+\bar{T}_{out}}{2}=T_0+\Delta T/2$), as noted from Fig.\ \ref{fig5}(c). Although both $\Delta T$ and $c_p$ decrease with increasing $\dot{m}$, increase in $\dot{m}$ leads to the increasing $\eta$ at initial stage. To be more specific, when $\dot{m}<0.06$ kg/s, the decreasing effects of $\Delta T$ and $c_p$ are less than the increasing effect of $\dot{m}$, so there is an obvious increase for $\eta$. For instance, for the coated DAPTSC, when $\dot{m}$ changes from 0.02 to 0.05 kg/s (increased to 250\%), $\Delta T$ decreases from 266.46 to 125.79 K (decreased to 47\%) and the corresponding decrease of $c_p$ is decreased from 2594 to 2361 J/kg-K (decreased to 91\%), resulting in $\eta$ increased to $107\% \approx 2.5 \times 0.47 \times 0.91$. For both DAPTSCs, when $\dot{m} > 0.1$ kg/s, the decreasing effect of $\Delta T$ and $c_p$ become nearly offset to the increasing effect of $\dot{m}$, leading to almost constant $\eta$. 

The second perspective for explaining the change of $\eta$ with $\dot{m}$ is through the energy balance analysis. Since $\dot{m}$ does not affect $A_m$, the solar energy absorbed in each collector is independent of $\dot{m}$. As a result, the change of $\eta$ depends mainly on the change of heat loss $Q_{loss}$; that is, the higher $Q_{loss}$ is, the lower $\eta$ will be. This is easily recognized in Figs.\ \ref{fig5}(a) and \ref{fig5}(d). Furthermore, the effect of $\dot{m}$ on $Q_{loss}$ can be understood as follows. At very low values of $\dot{m}$, the plasmonic nanofluid spends much time in the receiver tube; thus, higher $\Delta T$ can be gained as in Fig.\ \ref{fig5}(b). The higher $\Delta T$ in turn leads to the higher temperature on the receiver surface and consequently more $Q_{loss}$. When $\dot{m}$ increases, the corresponding $\Delta T$ decreases, and so does the $Q_{loss}$. The trend of $Q_{loss}$ is actually very similar to that of $\Delta T$ for the SBPTSC. While for the DAPTSCs, $Q_{loss}$ is much lower than that of the SBPTSC as the temperature on the inner tube surface is much lower than that of the SBPTSC owing to the volumetric absorption of the solar radiation.

It is also noted that $\eta$ of the DAPTSCs is higher than the SBPTSC until $\dot{m} = 0.1$ kg/s for the transparent DAPTSC and until $\dot{m} = 0.16$ kg/s for the coated DAPTSC. This is because when $\dot{m}> 0.2$ kg/s, the temperature gain $\Delta T$ for all the collectors are as low as about 34 K, leading to much less heat loss, especially for the SBPTSC. On the other hand, $A_m$ for the SBPTSC is 0.849; that is, much higher than 0.783 and 0.770 for the coated and transparent DAPTSC, respectively. Therefore, the SBPTSC gradually surpasses the transparent and coated DAPTSCs. Hence, the DAPTSCs are more desirable for low flow rate operation conditions (e.g., $\dot{m} < 0.16$ kg/s for the coated DAPTSC) while the SBPTSC works better for higher flow rate conditions (e.g., $\dot{m} > 0.18$ kg/s).

\subsection{Effects of the inlet temperature}
\begin{figure}[!t]
\centering
\includegraphics[width=0.5\textwidth]{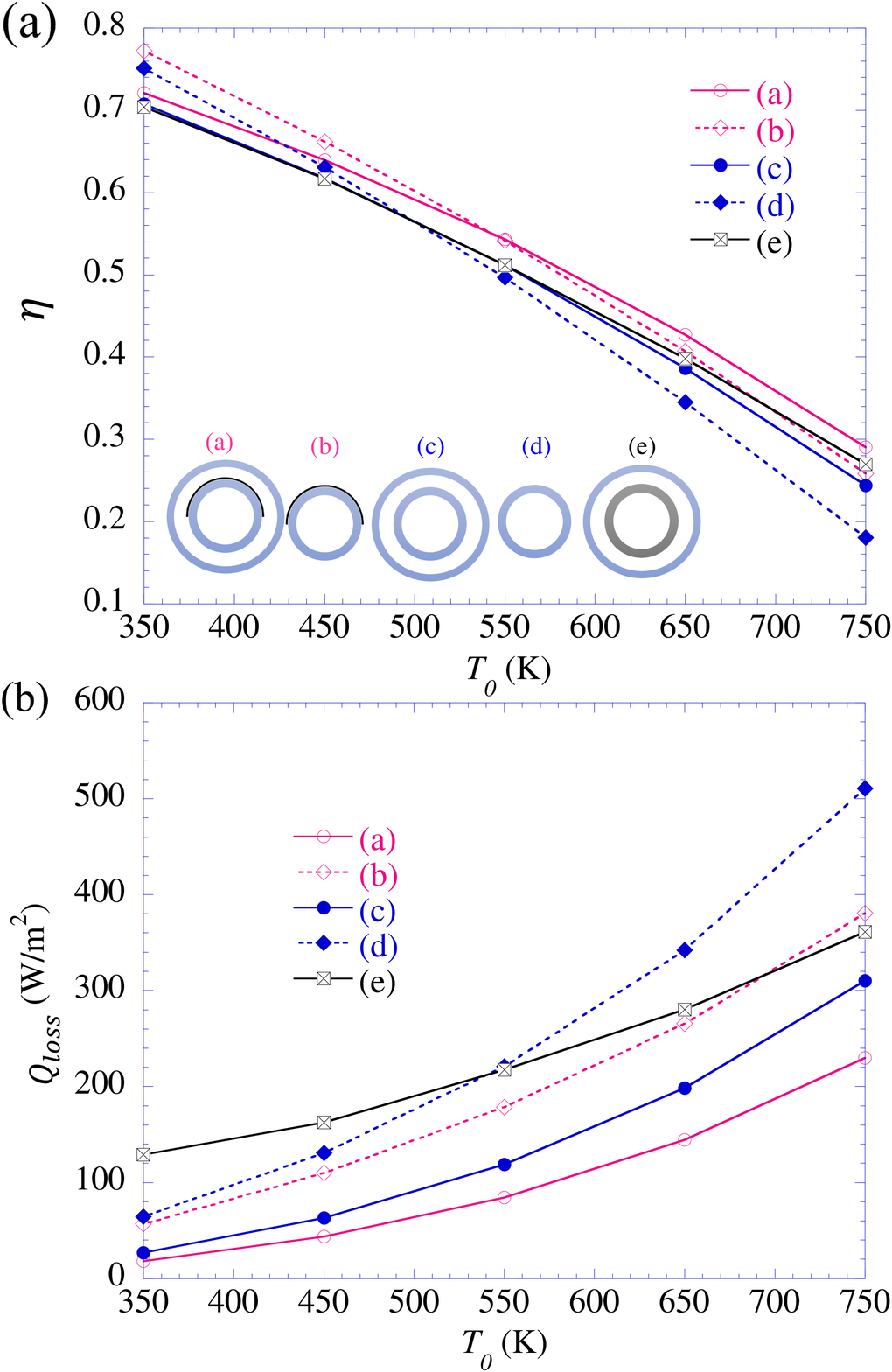}
\caption{Effects of the inlet temperature on (a) collector thermal efficiency and (b) heat loss.  In the calculation, we set $\alpha$ = 0.5 cm$^{-1}$, $D$ = 7 cm, $L$ = 5 m, and $\dot{m}$ = 0.1 kg/s.}
\label{fig6}
\end{figure}

In this section, effects of the inlet temperature on the thermal efficiency of five PTSCs (refer to Fig.\ \ref{fig2}) are studied and compared. The effect of the outer glass envelope for preventing the convection heat loss is also investigated for the two DAPTSCs (i.e., coated and transparent DAPTSCs). Note that the outer glass envelope is necessary for the SBPTSC for fair comparison due to its high surface temperature and thus high heat loss. First of all, it is seen from Fig.\ \ref{fig6}(a) that the collector thermal efficiency $\eta$ of all configurations decreases with increasing inlet temperature due to the increasing heat loss [see Fig.\ \ref{fig6}(b)]. Secondly, Fig.\ \ref{fig6}(a) also suggests that the best-performance configuration among the five collectors largely depends on the inlet temperature $T_0$. When $T_0 = 350$ K, for instance, the $\eta$ for the DAPTSCs are higher than the SBPTSC due to the high heat loss of the SBPTSC. Interestingly, the DAPTSCs without the glass envelop performs better than the corresponding DAPTSCs with the glass envelope. This indicates that the DAPTSCs without the glass envelope become more advantageous than the DAPTSCs with the glass envelope at low temperature owing to their higher $A_m$. In other words, at lower temperature applications, improving the solar weighted absorption coefficient by removing the glass envelope ($\because \tau_{gla}=0.94 < 1.0$) is more crucial than decreasing the convection heat loss. When $T_0 > 500$ K, however, the DAPTSCs with the glass envelope outperform the corresponding DAPTSC without the glass envelope. It is also confirmed again that the coated DAPTSCs are better than the transparent DAPTSC both with or without the envelope because the metallic coating doubles the optical path length and reduces the emissivity. As to the SBPTSC, its $\eta$ is lower than the nanofluid-based DAPTSCs at low temperature, but surpasses the configurations of (b), (c) and (d) sequentially at 510, 550 and 700 K. It is worth mention that $\eta$ of the transparent DAPTSCs with the envelope is very close to that of the SBPTSC when $350 < T_0 < 550$ K. On the other hand, the coated DAPTSC performs better than the SBPTSC until up to $T_0=750$ K mainly owing to the lower temperature on the inner tube surface than that of the SBPTSC.

\section{Conclusion}

In the present study, a partially coated direct-absorption parabolic trough solar collector utilizing the blended plasmonic nanofluid with uniform absorption coefficient has been proposed and investigated. In the proposed coated DAPTSC, the upper half of the inner tube's outer surface is coated by a reflective metallic layer so that the optical path length of the concentrated sun rays can be doubled as well as its infrared emissivity can be reduced significantly. Key factors influencing the collector performance have also been investigated. The following conclusions have been drawn. 

\begin{itemize}
	\item For a given inner tube's diameter of $D=7$ cm, the absorption coefficient of the blended plasmonic nanofluid directly affects the thermal performance of DAPTSCs, and higher $\alpha$ leads to higher $\eta$ until a certain value (i.e., $\alpha=0.4$ cm$^{-1}$ for the coated DAPTSC and $\alpha=0.8$ cm$^{-1}$ for the transparent DAPTSC). The proposed coated DAPTSC was found to be advantageous as compared to the transparent DAPTSC, especially at low $\alpha$ values (e.g., $\alpha < 0.5$ cm$^{-1}$). Therefore, to achieve the same level of $\eta$, the coated DAPTSC requires less nanoparticle concentration, which is beneficial for stable operation. 

	\item The collector thermal efficiency was found to be decreasing with increasing the diameter of the inner tube. We showed that there exists an optimal diameter $D$ for the DAPTSCs at which a reasonably high $\eta$ can be obtained with relatively low $\alpha$ if $\alpha D$ is fixed. 
		
	\item The DAPTSCs (both coated and transparent) were found to be more desirable when $\dot{m} \le 0.1$ kg/s, while the SBPTSC works better for the higher flow rate (e.g., $\dot{m} >  0.18$ kg/s). This is because when $\dot{m}> 0.2$ kg/s, the temperature gain $\Delta T$ for all the collectors are as low as about 34 K, resulting in much less heat loss, especially for the SBPTSC.
	
	\item  The DAPTSCs (both coated and transparent) with the glass envelope outperform the corresponding DAPTSC without the glass envelope until up to $T_0=500$ K for the transparent DAPTSC and up to $T_0=550$ K for the coated DAPTSC. It was found that for the low-temperature inlet conditions (e.g., $350 < T_0 < 430$ K), the transparent DAPTSC without the glass envelope is the best configuration. On the other hand, the coated DAPTSC with the glass envelope performs better than the SBPTSC until up to $T_0=750$ K. 
\end{itemize}

\noindent The results obtained in this work will facilitate the use of nanofluids in a parabolic trough solar collector with more stability and provide a guide for choosing suitable type of parabolic trough solar collectors for specific working conditions.

\section*{Acknowledgments}
This research was supported by Basic Science Research Program (NRF-2015R1A2A1A10055060) as well as Pioneer Research Center Program (NRF-2013M3C1A3063046) through the National Research Foundation of Korea (NRF) funded by the Ministry of Science and ICT.

\section*{References}

\bibliographystyle{elsarticle-num} 
\biboptions{sort&compress}


\bibliography{CSP_Qin.bib}

\end{document}